\documentstyle[12pt]{article}
\begin{document}
\bibliographystyle{unsrt}
\begin{center}
{\bf On the existence of localized excitations in
nonlinear Hamiltonian lattices}
\\[1cm]
S. Flach$^*$
\\[1cm]
{\sl Department of Physics, Boston University, 590 Commonwealth
Avenue, \\
Boston, Massachusetts 02215
}
\\[1.5cm]
\end{center}
$^*$
\small
present address: Max-Planck-Institut f\"ur
Physik Komplexer Systeme, Bayreuther Str.40 H.16, D-01187 Dresden, FRG.
\\
Email: flach@idefix.mpipks-dresden.mpg.de
\normalsize
\\
\\
\\
\\
ABSTRACT
\\
\\
We consider time-periodic nonlinear localized excitations (NLEs)
on one-dimensional translationally invariant Hamiltonian lattices
with arbitrary finite interaction range and arbitrary finite number
of degrees of freedom per unit cell.
We analyse a mapping of the Fourier coefficients of the NLE solution.
NLEs correspond to homoclinic points in the phase space of this map.
Using dimensionality properties of separatrix manifolds of the mapping
we show the persistence of
NLE solutions under perturbations of the system,
provided NLEs exist for the given system.
For a class of nonintegrable Fermi-Pasta-Ulam chains
we rigorously prove the existence of NLE solutions.
\vspace{0.5cm}
\\
PACS numbers: 03.20.+i, 63.20.Pw, 63.20.Ry
\\
{\sl Physical Review E}, in press.
\newpage
\section{Introduction}

The existence, stability and properties of nonlinear localized
excitations (NLEs) in Hamiltonian lattices with discrete
translational symmetry have been subjects of growing
research interest (see e.g.
\cite{st88},\cite{cp90},\cite{jbp90},\cite{st92} and references therein).
NLEs can be viewed as generalized discrete analogs to the breather
solution in a sine-Gordon equation \cite{cp90}. They are characterized
by a localized vibrational state of the lattice.
There are two basic reasons for the generic
NLE existence on Hamiltonian lattices: i) the lattice forces
acting on a given particle are { \sl nonlinear} (thus one
can tune oscillation frequencies by varying the energy) and
ii) the {\sl discrete translational symmetry} of the lattice
(in contrast to the continuous translational symmetry of Hamiltonian
field equations) provides a {\sl finite} upper phonon band edge
of the spectrum of extended small amplitude oscillations of the
lattice around its groundstate \cite{fw2},\cite{fw7}.

There are a few known rigorous NLE existence proofs.
First NLEs are exact solutions of
the {\sl integrable} Ablowitz-Ladik lattice
\cite{al76}.
In fact they form a three-parameter family of solutions.
Secondly
NLEs are exact solutions for
the Fermi-Pasta-Ulam chain with box-like interaction potential \cite{sps92}.
In this case the NLEs are of compact support.
Finally, and most importantly, MacKay and Aubry have derived
an existence proof for NLEs in an array of weakly coupled
anharmonic oscillators \cite{ma94}. Remarkably this existence proof works
independent of the lattice dimension.

In this contribution we will first deal with the existence of NLEs
in {\sl nonintegrable} generic one-dimensional Hamiltonian
lattices. In the second part we will investigate a class of
Fermi-Pasta-Ulam chains and give rigorous proofs for the
NLE existence.

Let us briefly outline the main steps in the approach presented
below. We will assume the existence of a time-periodic NLE
on a one-dimensional lattice. We represent
the lattice displacements at each lattice site in the NLE ansatz
in a Fourier series with respect to time. When we insert the NLE ansatz
into the lattice equations of motion, we obtain a set of coupled
algebraic equations for the Fourier components of the NLE ansatz,
which form an infinite-dimensional map. The NLE solution has to correspond
to a common point of two separatrix manifolds in the phase space
of the map, or a homoclinic point.
Analyzing the map in the (linearizable) tails of the NLE,
we can derive the dimension of the separatrix manifolds. Consequently
we show that if a homoclinic point exists for a given system,
then generically the homoclinic point will survive
under perturbations of the system. We then consider
a subclass of Hamiltonian chains and rigorously prove the existence
of two different (with respect to symmetry) NLE solutions.
For this particular example we show the emergence of horseshoe patterns
- a consequence of the existence of homoclinic points.

\section{Stability of NLE solutions under Hamiltonian perturbations}

We consider a classical one-dimensional Hamiltonian lattice
of interacting particles (perhaps feeling an external
field periodic with the lattice) with lattice cite $a=1$.
The displacements of the particles from their groundstate
(equilibrium) positions are given by a $n$-dimensional
vector $\vec{X}_l$, where $n$ is the number of components
per unit cell ($n \leq n_0$, $n_0$ finite) and the integer
$l$ marks the number of the unit cell. The range of the interaction
$r$ is considered to be finite: $r \leq r_0$, $r_0$ finite.
Here $n_0$ and $r_0$ are positive integers.
The potential energy of the system is required to have anharmonic
terms in the displacements if expanded in a Taylor series around
the groundstate (minimum of potential energy) of the system.
Furthermore the potential energy should become a positive definite
quadratic form in the limit of infinitely small displacements.
The Hamilton function $H$ is given by the sum of the kinetic
energy of all particles and the potential energy.

As it was shown in \cite{fw2},\cite{fw7},
the only possible { \sl exact} NLE solution
on an arbitrary lattice (by that we mean no additional
symmetries are present) has to have the form:
\begin{equation}
\vec{X}_l(t)=\vec{X}_l(t+2\pi/\omega_1)\;\;,\;\;
\vec{X}_{l \rightarrow \pm \infty} \rightarrow 0 \;\;. \label{1}
\end{equation}
Then one can avoid resonance conditions of multiples of the
fundamental frequency $\omega_1$ (as they appear in the Fourier
transformed functions in (\ref{1}) with respect to time because
of the nonlinearity of the system) with phonon frequencies
of the linearized (around the groundstate) system. Since the
motion of assumed existent NLEs requires the
excitation of at least a second fundamental frequency
in the ansatz (\ref{1}) \cite{fw6}, we can exclude them from the consideration
and search for {stationary time-periodic} NLEs as given in (\ref{1}).
Note that in the case of the Ablowitz-Ladik lattice additional symmetries
are present (the lattice is integrable) and thus the above statement
does not hold - moving NLEs are exact solutions in this
nongeneric case.

Because of the assumed time-periodicity of all displacements in (\ref{1})
we obtain for the assumed NLE solution
\begin{equation}
\vec{X}_l = \sum_{k=0,\pm 1, \pm 2, ...} \vec{A}_{(l,k)}{\rm e}
^{i\omega_1 k t}\;\;. \label{2}
\end{equation}
Now we can insert this ansatz (\ref{2}) into the Newtonian equations
of motion $\ddot{\vec{X}}_l = \partial H / \partial \vec{X}_l$.
The left and right hand expressions of the equation of motion
are represented again as a general Fourier series.
Equaling the prefactors at identical exponential terms to each other
we finally obtain a coupled set of algebraic equations for
the unknown Fourier coefficients $\vec{A}_{(l,k)}$. Because of (\ref{1})
the Fourier coefficients have to satisfy the boundary condition
$\vec{A}_{(l \rightarrow \pm \infty,k)} \rightarrow 0$.
In the following we will study properties of this algebraic set
of equations.

First we can note that because of the $d=1$ dimensionality of the
considered lattice the coupled set of algebraic equations for
the Fourier coefficients $\vec{A}_{(l,k)}$ can be represented
as a discrete map of a $d_M=(2r_0n_0k_{max})$-dimensional phase space,
where the integer $k_{max}$ represents the number of
considered higher harmonics in (\ref{2}) and has to tend
to infinity.
In other words, given the Fourier coefficients
at $2r_0$ neighboring lattice sites completely determines
the Fourier coefficients to the left and right of the specified
chain segment.
Secondly because of the required asymptotic vanishing of the Fourier
coefficients for $l \rightarrow \pm \infty$ we can linearize
the $d_M$-dimensional map in the tails of the assumed NLE solution
with respect to the variables $\vec{A}_{(l,k)}$.
The linearized map will decouple into a set of $k_{max}$ independent
$d_s=2r_0n_0$-dimensional linear submaps \cite{fw7}. In each of these submaps
Fourier components with only one Fourier integer $k$ will appear.
Each linear submap is equivalent to the problem of finding
solutions of the linearized (around the groundstate) lattice equations
of motion using the ansatz $\vec{X}_l(t)= A_l \exp{i\tilde{\omega}t}$.
Here for every submap one has to substitute $\tilde{\omega}= k\omega_1$.
If $\tilde{\omega}$ equals to an arbitrary phonon frequency of
the linearized lattice equations, then it follows that the
corresponding linear submap is characterized by a matrix ${\bf M}_k$
of dimension $d_s \times d_s$ with ${\rm det}{\bf M}_k=1$ and
$\lambda_i \lambda_{i+d_s} =1$ where $\lambda_{i,i+d_s}$ are two
of the $2d_s$ eigenvalues
of ${\bf M}_k$. Since these properties do not depend on the specific
value of the considered phonon frequency, it follows that they
are independent on $\tilde{\omega}$ and thus independent on $k$
and $\omega_1$. Consequently we find that every linear submap
is volume preserving and symplectic.

Finally we note that for every considered linear submap (and thus
also for the original $d_M$-dimensional map) the phase space point of
zero Fourier components is a fixed point of the map.
If $k\omega_1$ equals a phonon frequency, the fixed point is
of elliptic character. If however $k \omega_1$ does not equal
any phonon frequency, the fixed point has to be a saddle point
(because it can not be an elliptic fixed point and the map is
symplectic). That means that $r_0n_0$ eigenvalues of the matrix ${\bf M}_k$
are real and of absolute value lower than one. Consequently
the separatrix manifold of all points in the space of the
submap attracted by the saddle point is of dimension $d_{sm}=r_0n_0$.

If we require that neither of the multiples of the fundamental
frequency $k\omega_1$ resonates with the phonon band, the
corresponding original $d_M$-dimensional map in the phase space
of the Fourier coefficients has a separatrix manifold
of dimension $d_M/2$. To get a NLE solution
we have to find a point in the space of the original map which
belongs simultaneously to the separatrix manifold $S_-$ attracting the
solution to zero for $l \rightarrow - \infty$ and to the
separatrix manifold $S_+$ attracting the solution
to zero for $l \rightarrow +\infty$ (cf e.g. \cite{jd93},\cite{ekns84}).
This point has to be then a homoclinic point \cite{ll92}. If a homoclinic
point exists, there exist an infinite number of homoclinic points, which
can be obtained by subsequent mapping of a given homoclinic point.
As a result the horseshoe structure of intersections between the
stable and unstable manifolds have to emerge \cite{ll92}.

Because of the assumed nonresonance condition (see above) every manifold
has dimension $d_M/2$. Let us choose one of the homoclinic points.
Then we can define the tangent planes to each of the two manifolds
in this point.
There are two possibilities for the topology of these two planes.
Either i) these two planes span the whole phase space
of the map (of dimension $d_M$) or ii) the two planes span a
space of lower dimension than $d_M$. In case i) any small perturbation
of the original system (consequently of the map, consequently
of the two manifolds and consequently of the two tangent planes)
will only shift the homoclinic point smoothly. Also in this case i)
all other homoclinic points will have the same topology with
respect to the tangent planes. In case ii) there exist perturbations
of the system which will lead to a vanishing of the homoclinic point
(and thus of all other homoclinic points too). Still there will
exist a perturbation for case ii) such that the homoclinic point
is smoothly shifted {\sl and} simultaneously the two tangent maps
will span the whole phase space of dimension $d_M$.

Consequently we can conclude, that if we have a NLE solution which
corresponds to a set of homoclinic points of type i) then any small
perturbation of the system will smoothly transform
the NLE solution into a new NLE solution.
If we have a NLE solution which corresponds to a set of homoclinic
points of type ii), we can always find a perturbation such that
we transform the NLE solution into a new NLE solution which corresponds
to a set of homoclinic points of type i) and is then stable under
subsequent small perturbations. From the above said it follows
that NLE solutions, if they appear, are generically not isolated
objects in the sense that small perturbations of the system
either do not destroy them at all, or that a proper perturbation
of the system will transform them into nonisolated solutions.

We have operated with finite phase space dimensions
$d_M$. Of course $d_M$ was not bounded from above, so that the
limit $d_M \rightarrow \infty$ can be considered without altering
the arguments.

Let us summarize the results obtained so far. If we assume that
for a given Hamiltonian chain a NLE solution exists, then it
corresponds to an infinite number of homoclinic points in the
map as introduced above. If neither of the multiples of the
frequency of the NLE solution resonate with the phonon band
(nonresonance condition),
then the dimension of the stable and unstable manifolds is
exactly one half of the map's phase space. From topological
arguments it follows then, that either the NLE solution is structurally
stable - i.e. it is smoothly transformed under {\sl any} perturbation
of the Hamiltonian of the system, or there exists at least one perturbation
of the Hamiltonian such, that the NLE solution is smoothly transformed
into a structural stable one. Of course a perturbation of the Hamiltonian
has to preserve the general structure of the map, i.e. we are
not allowed to consider e.g. a two-dimensional perturbation, which
would lead to the impossibility of defining the map.

A central point in the consideration so far has been the nonresonance
condition. If this condition is violated for $n$ values of
$k\omega_1$, then the dimension of the stable and unstable manifolds
is lowered by $(2n)$ to $(d_M - 2n)$. If an NLE solution exists
for such a case, then the tangent planes of the two manifolds
in a given homoclinic point can not span the whole phase space
of the map. Consequently there will be always perturbations of
the Hamiltonian such that the NLE solution will vanish.
At this point it also becomes clear that in the analogous problem
of a Hamiltonian field equation, where resonances can be never
avoided \cite{ekns84}, systems which allow for NLE solutions
become isolated (see also \cite{jd93}).

If a system is given with a NLE solution which fulfills the
nonresonance condition, it could be possible that a given
perturbation of the Hamiltonian would lead to a violation
of the nonresonance condition. Consequently one has to check in
a given case, whether the nonresonance condition survives under
a perturbation.

\section{Proof of existence of NLE solutions for Fermi-Pasta-Ulam
chains}

In the last part of this work we will prove rigorously the
existence of NLE solutions for a class of Fermi-Pasta-Ulam
chains governed by the following equations of motion:
\begin{equation}
\ddot{X}_l=-(X_l-X_{l-1})^{2m-1}-(X_l-X_{l+1})^{2m-1}\ \label{3}
\end{equation}
with $m=2,3,4,...$.
As it was shown in \cite{fw7},\cite{ysk93} we can consider a time-space
separation ansatz $X_l(t)=A_l G(t)$. Inserting the separation
ansatz into (\ref{3}) we get a differential equation for $G(t)$:
$\ddot{G}(t)=-\kappa G^{2m-1}(t)$ and a two-dimensional map
for the amplitudes $A_l$:
\begin{equation}
\kappa A_l= (A_l-A_{l-1})^{2m-1}+(A_l-A_{l+1})^{2m-1}\;.\label{4}
\end{equation}
Here $\kappa > 0$ is required in order to get a bound oscillatory
solution for $G(t)$. The phase space properties of map (\ref{4})
are shown in Fig.1 for the case $m=2$ and $\kappa=1$. Below we
will refer to particular patterns observed in Fig.1.
We consider cases when $A_l/A_{l-1} < 0$
and introduce $f_l=|-1|^{l}A_l$ with
\begin{equation}
\kappa f_l = (f_l+f_{l-1})^{2m-1}+(f_l + f_{l+1})^{2m-1}\;. \label{5}
\end{equation}
Equation (\ref{5}) can be viewed as a two-dimensional map $\bf M$
of a vector $\vec{f}_l=(f_l,f_{l-1})$: $\vec{f}_{l+1}={\bf M} \vec{f}_l$.
The
task is then to show the existence of at least one
value of $\kappa > 0$ such that (\ref{5})
yields $f_{|l|} \geq f_{|l'|} > 0$ for $|l| < |l'|$ and $f_{l \rightarrow
\pm \infty} \rightarrow 0$.

First we note that for any value of $\kappa$ the map (\ref{5})
has a fixed point $\vec{f}_{fp1}=(0,0)$. Adding weak harmonic
nearest neighbour interactions to (\ref{3}) and considering
the limit of vanishing harmonic interactions yields that the
fixed point $(0,0)$ is a saddle point with eigenvalues
$\lambda_{1,fp1}=0$ and $\lambda_{2,fp1}= \infty$.
Consequently we find that there exist two one-dimensional
separatrix manifolds $S_+$ and $S_-$ of the map (\ref{5}).
All points in the two-dimensional phase space of the map
belonging to $S_{\pm}$ are attracted to the saddle point
after an infinite number of iterations for $l \rightarrow \pm \infty$.
Secondly
it follows that for $\kappa_{fp2}=2^{2m}f^{2m-2}$
the point $\vec{f}_{fp2}=(f,f)$ is also a fixed point (see also Fig.1).
Linearizing the map around
$\vec{f}_{fp2}$ yields the elliptic character of $\vec{f}_{fp2}$.
The two eigenvalues of the linearized map $\lambda_{1,fp2}$ and
$\lambda_{2,fp2}$ obey the relations $|\lambda_{1,fp2}|=|\lambda_{2,fp2}|=1$
and are given by the expression $\lambda_{12,fp2}=\tilde{\kappa}-1
\pm i (1-(1-\tilde{\kappa})^2)^{1/2}$ with $\tilde{\kappa}=2/(2m-1)$.

Let us mention another useful property of the map (\ref{5}),
which holds due to the discrete translational symmetry
of the considered system (\ref{3}).
If we generate a sequence of $f_2,f_3,...$ starting with a vector
$\vec{f}_1=(a,b)$ and iterating (\ref{5}) towards growing integers $l$,
then we can generate the same sequence of numbers
by iterating (\ref{5}) towards decreasing integers $l$ starting
with the vector $\vec{f}_1=(b,a)$.

We consider an initial vector $\vec{f}_l=(a,b)$ with $0 < a \leq b$
and iterate towards increasing $l$. If we require $0 \leq f_{l+1} \leq a$,
the parameter $\kappa$ is confined to $\kappa_l^- \leq \kappa
\leq \kappa_l^+$
with $\kappa_l^- = ((f_l+f_{l-1})^{2m-1} + f_l^{2m-1})/f_l$ and $\kappa_l^+
= \kappa_l^- + (2^{2m-1}-1)f_l^{2m-2}$. For any of the obtained values
of $f_{l+1}=c$ we can consider the next step requiring $0 \leq f_{l+2}
\leq c$. By itself this requirement again yields a confinement
of $\kappa$ to $\kappa_{l+1}^- \leq \kappa \leq \kappa_{l+1}^+$.
If we choose $c=0$ then $\kappa_{l+1}^- = +\infty$. If we choose
$c=a$ it follows $\kappa_{l+1}^+ \leq \kappa_l^+$. Because $c$ is
a monotonic function of $\kappa$ we can satisfy $a \geq f_{l+1} \geq
f_{l+2} \geq 0$ only if $\kappa_{l}^- < \kappa \leq \kappa_l^+$ for
$a=b$ and $\kappa_l^- < \kappa < \kappa_l^+$ for $a < b$.
Repeating this procedure for every following lattice site
we get a sequence of monotonically increasing lower bounds on
$\kappa$ defined by the
vanishing of $f_{l+m}=0$ in the $m$-th step. In the limit $m \rightarrow
\infty$ we thus obtain exactly one value of $\kappa$, which is
the limit of the mentioned sequence of lower bounds. This limiting
value is finite, because it is always smaller than $\kappa_l^+$.
It can not be equal to $\kappa_l^+$ either, since we would
then get the infinitely weak perturbed elliptic point of the
map. Because of the elliptic character of this fixed point
all amplitudes $f_l$ have to stay infinitely close to their
fixed point values which are finite. Thus we have found, that
for any initial vector $\vec{f}_l=(a,b)$ with $a \leq b$ there exists
exactly one value of $\kappa$ such that the initial vector
belongs to $S_+$. The point $\vec{f}_l=(b,a)$ belongs then
to $S_-$.

In the case $a=b$ it follows that it is always possible
to find exactly one value of $\kappa$ such that the initial
vector $\vec{f}_l=(a,a)$ belongs to {\sl both} $S_+$ and $S_-$.
Consequently we proved rigorously the existence of one type
of NLEs, which are known as 'even-parity-mode' \cite{jbp90}.
Their characteristic
feature is that the center of energy density of the corresponding
solution is located between two lattice sites $l$ and $(l-1)$ at
$(l-0.5)$ \cite{fw6}.

Finally let us consider the case $0 \leq f_{l-2}=f_{l}=a \leq f_{l-1}=b $.
Solving the equation (\ref{5}) for $l \rightarrow (l-1)$ we get
allowed values of $\kappa$ in an interval defined by $b$. Appending
the above described procedure for the sequence of lower bounds
of $\kappa$ we again find that for exactly one value of $\kappa$
such that $a < b$ the vectors $(a,b)$ and $(b,a)$ belong
to $S_+$. By the symmetry of the assumed initial configuration they
also belong to $S_-$. Consequently  we proved rigorously the
existence of a second type of NLEs, which are known as 'odd-parity mode'
\cite{jbp90}.
Their characteristic feature is that the center of energy density of
the solution is located on the lattice site $(l-1)$ \cite{fw6}.

Due to the radius of interaction $r_0=1$ it follows that there
are no other allowed stationary time-periodic NLE solutions
in lattices of type (\ref{3}). In the limit $m \rightarrow \infty$
the two NLE solutions become compact and can be calulated
analytically \cite{sps92}.

It would be now very interesting to analyse the map (\ref{5})
in order to observe the previously discussed stable and unstable
manifolds, and consequently the homoclinic points. However
this map (\ref{5}) is not linearizable around the fixed point $f_l=0$ -
some matrix elements of the Jacobian diverge. Consequently
it is impossible to determine the structure of the manifolds
near the fixed point numerically. In order to proceed we
add to the right hand side of (\ref{5}) the terms
\begin{equation}
 C(f_l+f_{l-1}) + C(f_l + f_{l+1}) \;\;. \label{10}
\end{equation}
In this case the modified map becomes linearizable around
the fixed point $f_l=0$. We can then visualize the manifolds,
and hope that in the limit $C \rightarrow 0$ the structure
of the manifolds does not change with respect to their intersections.
In Fig.2 we show the stable and unstable manifolds for
$m=2$, $C=2$ and $\kappa = 10$, where we used the variables $g_l=f_l+f_{l-1}$
instead of the original ones.
We computed the unstable manifold using standard procedures \cite{ll92}
and indicated the position of the stable manifold as it can be
evaluated out of the linearization around the fixed point.
Clearly the horseshoe patterns
are visible together with the homoclinic points. If one chooses one
homoclinic point and iterates, then the image is the next-to-next
homoclinic point. Thus we indeed observe the two possible
even and odd parity solutions. If we lower the value of $C$, then
it becomes increasingly harder to find the unstable manifold.
Still the structure of the unstable manifold can be seen partially, which
indicates the persistence of the horseshoe patterns in phase space regions
where the linear terms (\ref{10}) are small perturbations.
Then we can use the results of Section II and conclude that NLE solutions
will survive under perturbations of the considered system. In particular
it follows that NLEs are exact solutions for Fermi-Pasta-Ulam chains
with additional linear and cubic spring forces, as often studied
in the literature \cite{st88},\cite{st92}.

\section{Conclusion}

In conclusion we have shown, that if a one-dimensional
system is known where
NLE solutions exist and obey the nonresonance condition, then
they are stable under perturbations of the Hamiltonian of the
system or they can be perturbed in such a way that they
become stable. Consequently we can use exact proofs of the
existence of NLE solutions, e.g. the one given in this work
for (\ref{3}) or the one in \cite{ma94} and obtain new NLE solutions
for the corresponding perturbed system. Remarkably we are not
restricted in the choice of the perturbation as long as the
map for the perturbed system has the same phase space dimension.

Let us discuss the problems of the presented approach which
appear when we try to consider lattices of dimensionality higher
than $d=1$. We can still make the ansatz of an existing time-periodic
NLE solution and will yield again a coupled set of algebraic
equations for the Fourier coefficients. However this set can not
be viewed as a discrete map of a certain phase space of the
Fourier components, if $d > 1$. Consequently we do not know
how to formulate the conditions of the decay of the solution
at infinity by imposing certain constraints on the set of equations.
Still it is well known from numerical simulations that periodic
NLEs exist \cite{fw8}.
\\
\\
\\
Acknowledgements
\\
\\
I wish to thank C. R. Willis for his continuous support of
these studies and for the many helpful discussions. I
also thank M. Peyrard, D. K. Campbell, K. Kladko,
U. Bahr and B. Flach for
interesting discussions and comments and S. Aubry for
sending a preprint prior publication. This work was supported
by the Deutsche Forschungsgemeinschaft (Fl 200/1-1).

\newpage

FUGURE CAPTIONS
\\
\\
\\
Fig.1
\\
\\
Phase space of the map defined by (\ref{4}) for $\kappa=10$ and $m=2$.
The fixed points are: $\vec{A}_{fp1}= (0,0)$ (solid circle),
$\vec{A}_{fp2}=(\pm x, \mp x)$ ($x=1/\sqrt{1.6}$, periodicity 2),
$\vec{A}_{fp3}=(0,\pm y)$ and $(\pm y,0)$ ($y=\sqrt{5}$, periodicity 4).
\\
\\
\\
\\
Fig.2
\\
\\
Appearance of the homoclinic points and horseshoe structure
for the perturbed map (\ref{5}),(\ref{10})
with variables as given in the text.

\newpage


\begin{thebibliography}{10}

\bibitem{st88}
A.~J. Sievers and S.~Takeno.
\newblock Phys. Rev. Lett.~61, 970 (1988).

\bibitem{cp90}
D.~K. Campbell and M.~Peyrard.
\newblock {\em in: CHAOS - Soviet American Perspectives on Nonlinear Science,
  ed. by D. K. Campbell}.
\newblock American Institute of Physics New York, 1990.

\bibitem{jbp90}
J.~B. Page.
\newblock Phys. Rev.~B41, 7835 (1990).

\bibitem{st92}
S.~Takeno.
\newblock J. Phys. Soc. Japan~61, 2821 (1992).

\bibitem{fw2}
S.~Flach, C.~R. Willis, and E.~Olbrich.
\newblock Phys. Rev. E~49, 836 (1994).

\bibitem{fw7}
S.~Flach.
\newblock Phys. Rev. E~, in press (October 1994).

\bibitem{al76}
M.~J. Ablowitz and J.~F. Ladik.
\newblock J. Math. Phys.~17, 1011 (1976).

\bibitem{sps92}
K.~W. Sandusky, J.~B. Page, and K.~E. Schmidt.
\newblock Phys. Rev.~B46, 6161 (1992).

\bibitem{ma94}
R.~S. MacKay and S.~Aubry.
\newblock Warwick Report No. 20/1994~, unpublished (1994).

\bibitem{fw6}
S.~Flach and C.~R. Willis.
\newblock Phys. Rev. Lett.~72, 1777 (1994).

\bibitem{jd93}
J.~Denzler.
\newblock Commun. Math. Phys.~158, 397 (1993).

\bibitem{ekns84}
V.~M. Eleonskii, N.~E. Kulagin, N.~S. Novozhilova, and V.~P. Shilin.
\newblock Teor. Mat. Fiz.~60, 395 (1984).

\bibitem{ll92}
A.~J. Lichtenberg and M.~A. Lieberman.
\newblock {\em Regular and Stochastic Motion}.
\newblock Springer, New York, 1992.

\bibitem{ysk93}
Yu.~S. Kivshar.
\newblock Phys. Rev. E~48, R43 (1993).

\bibitem{fw8}
S.~Flach, K.~Kladko, and C.~R. Willis.
\newblock Phys. Rev. E~, in press (September 1994).

\end{thebibliography}
\end{document}